\documentclass[intlimits,twoside,a4paper]{article}

\usepackage{graphicx}
\usepackage[cp1251]{inputenc}

\usepackage{cmpj3}

\usepackage{amssymb}
\usepackage{amsmath}

\issue{2019}{22}{1}{13602}
\doinumber{10.5488/CMP.22.13602}

\title[Water-methanol mixtures]
{Molecular dynamics simulations of the properties of water-methanol mixtures. 
Effects of force fields} 

\author[M. Cruz Sanchez, H. Dominguez,
O. Pizio]{M. Cruz Sanchez\refaddr{label1}, H. Dominguez\refaddr{label2},
O. Pizio\refaddr{label1} \thanks{Corresponding author, E-mail: oapizio@gmail.com}}

\addresses{
\addr{label1} Instituto de Qu\'{i}mica, Universidad Nacional Aut\'{o}noma de M\'{e}xico,
Circuito Exterior, \\ 04510 Cd. de M\'{e}xico, M\'{e}xico
\addr{label2} Instituto de Investigaciones en Materiales, Universidad Nacional Aut\'{o}noma de M\'{e}xico,
Circuito Exterior, \\ 04510 Cd. de M\'{e}xico, M\'{e}xico}

\date{Received February 22, 2019}

\begin{document}
\maketitle

\begin{abstract}
Isothermal-isobaric  molecular dynamics simulations 
are used to examine the microscopic structure and some properties
of water-methanol liquid mixture. 
The TIP4P/2005 and SPC/E water models are combined 
with the united atom TraPPE and the all-atom force field model for methanol. 
Our principal focus is to evaluate the quality of predictions of
different combinations of model force fields concerning the composition dependence of 
basic properties of this system.
Specifically, we explored the composition effects on density, excess
molar volume and excess entropy, as well as on the surface tension and static
dielectric constant. In addition, the
structural properties are described in terms of the coordination numbers 
and the average number of hydrogen bonds between molecules of constituent species.
Finally, the composition dependence of self-diffusion coefficients of the species
is evaluated. All theoretical predictions are tested with respect to experimental data.


\pacs 61.20.-p, 61.20-Gy, 61.20.Ja  

\end{abstract}

\section{Introduction}

This manuscript is the first part of our two-stage project that involves water-methanol
mixtures. Namely, in this first part, we present a set of results coming from the 
isothermal-isobaric ($NPT$)
computer simulation concerning composition changes of density, excess mixing volume and
entropy, first coordination numbers of species and average number of hydrogen bonds.
Moreover, we explore the behavior of the surface tension and the dielectric constant  
on composition, as well as the self-diffusion coefficients of species.

The forthcoming, second part of the project, is devoted to the exploration of changes
of all the properties mentioned above, brought by addition of NaCl salt to water-methanol
solvent. Again, the $NPT$ computer simulation technique will be applied. One of
the principal issues we would like to address is to evaluate the validity and quality of 
theoretical results coming from different combinations of force fields with respect 
to experimental data.  

For specific purposes of our project, it is worth mentioning that molecular dynamics computer
simulations have been widely applied to mixtures of water with various solvents.
Most frequently studied seem to be the mixtures of water with alcohols, specifically
the water-methanol mixtures,
see, e.g.,~\cite{wensink,galicia1,galicia2,perera1,perera2,bako,bopp1,bopp2,jordan,matsumoto,chang} 
and references therein for a rather comprehensive account of the previously applied modelling. 
Our present report has been inspired by most recent contributions
concerning the system in question~\cite{bako,bopp2,jordan}.
On the other hand, important experimental observations concerning this system and used 
in the present study have been discussed in~\cite{takamuku,wakisaka,mikhail,mcglashan,lama,vazquez}.

The majority of computer simulation studies cited above have been focused on the application
of a single combination of force fields describing each constituent species, water and
methanol with the exception of~\cite{jordan}. This latter work involved two methanol
force fields, namely the TraPPE model~\cite{trappe}  and OPLS-all atom model~\cite{jorgensen2}.  
As concerns water, the SPC/E and TIP4P models were chosen, see, e.g.,~\cite{spce} and~\cite{jorgensen3},
respectively.  Formally, we apply similar strategy. However, in contrast to~\cite{jordan}, 
we apply the TIP4P/2005 water model~\cite{vega-tip} rather than the TIP4P version~\cite{jorgensen3}.
It is known that the the TIP4P/2005 version provides a better performance of the
microscopic structure of water~\cite{pusztai}  and yields even better description of various 
properties compared to the TIP4P, see, e.g.,~table 2 of reference~\cite{vega-pccp}.
Moreover, we are specifically interested in using the TIP4P/2005, because it has been quite recently applied to parametrize the description of properties of NaCl aqueous solution, 
see~\cite{benavides1,madrid}. It opens up the possibility to study NaCl solutions with
water-methanol solvent in subsequent work. On the other hand, the gained
knowledge would permit to explore complex solutes in mixed solvents tuning with 
confidence their solubility on solvent composition. 
Some specific molecules of interest in medicinal chemistry, for example curcumin,  
are marginally soluble in water, but dissolve well in, e.g., alcohols or 
dimethylsulfoxide~\cite{bagchi,roccatano,patsahan}.
Additional comments concerning the
methodological difference of our procedure and technical details in comparison with reference~\cite{jordan}
are given in the body of the manuscript below.

To summarize, the principal objective of the present report
is to investigate a set of properties of water-methanol mixtures in the entire  interval 
of composition. All the properties are validated by comparison with experimental results.

\section{Models and simulation details}

Preliminarily discussing the simulation methodology, for the sake of convenience for the reader, in table~\ref{tab:table1}
we list some of the models used in previous studies of water-methanol mixtures and 
related to the present work. Within the united atom model for
methanol, like OPLS/UA and TraPPE (see table~\ref{tab:table1}), the CH$_3$ group is considered as a single
site. On the other hand, the all atom models explicitly involve all the hydrogens of the methanol molecule.
\begin{table}[h!]
    \caption{Models of water-methanol liquid mixtures and the combination rules (according to
     GROMACS nomenclature).}
    \label{tab:table1}

    \vspace{0.1cm}
      \begin{center}
        \begin{tabular}{l |l |l |l }

     \hline
         & water & methanol& combination rules \\
    \hline
     Galicia-Andres \textsl{et al.}~\cite{galicia2} & TIP4P/Ew~\cite{horn}  & OPLS/UA~\cite{jorgensen,haughney}&  CR2 (L-B)\\
                                           & SPC/E~\cite{spce}    &                                   &         \\
\hline
     Galicia-Andres \textsl{et al.}.~\cite{galicia1} & TIP4P/Ew ~\cite{horn} & OPLS/AA~\cite{jorgensen2} &      CR3 \\
                                           & SPC/E~\cite{spce}     &            \\
\hline
  Wensink \textsl{et al.}~\cite{wensink}  & TIP4P~\cite{jorgensen3}  & OPLS/AA~\cite{jorgensen2} &     CR3 \\                               
     \hline
    Guevara-Carrion \textsl{et al.}~\cite{guevara} &  SPC/E~\cite{spce} & UA-own design~\cite{vrabec2} &        CR2 (L-B)\\
                                          & TIP4P/2005~\cite{vega-tip}    &         & \\
     \hline
     Kohns \textsl{et al.}.~\cite{hasse1} &  SPC/E~\cite{spce} & UA-own design~\cite{vrabec2} &        CR2 (L-B)\\
     \hline

       Po$\check{z}$ar \textsl{et al.}~\cite{pozar} &  SPC/E~\cite{spce} & TraPPE~\cite{trappe} &   CR2 (L-B)\\
     \hline
     Obeidat \textsl{et al.}~\cite{jordan} & TIP4P~\cite{jorgensen3}  & TraPPE~\cite{trappe} &  CR2 (L-B)\\
                                  & SPC/E~\cite{spce}    &     OPLS/AA~\cite{jorgensen2}        &         \\
\hline
\end{tabular}
\end{center}
\end{table}

\subsection{Technical details}

In this work, we explore the SPC/E model~\cite{spce} and the 
TIP4P/2005 model~\cite{vega-tip} for water.
For methanol, we used two models, namely
the united atom model-TraPPE~\cite{trappe},
and all atom model~\cite{jorgensen2}, denominated as  MET/TraPPE, and MET/AA, respectively.
The Lorentz-Berthelot combination rules are used to determine the cross parameters.

The long-range electrostatic interactions were handled by the
particle mesh Ewald method implemented in the GROMACS software package (fourth
order, Fourier spacing equal to 0.12) with a precision of $10^{-5}$.
The nonbonded interactions were cut off at 1.4 nm.
The van der Waals tail correction terms to the energy and pressure were taken into account.
In order to maintain the geometry of the water and methanol  molecules, the LINCS
algorithm was used.

Our calculations were performed in the isothermal-isobaric ($NPT$) ensemble
at 1~bar, and at a temperature of 298.15~K. We used the GROMACS
software package~\cite{gromacs}, version 5.1.2.
Concerning the procedure, a periodic cubic simulation box was set up for each system. 
The GROMACS genbox tool was employed to randomly place all particles in the 
simulation box.
To remove the possible overlaps of particles introduced by the procedure of
preparation
of the initial configuration, each system underwent energy
minimization using the steepest descent algorithm implemented in the GROMACS
package. Minimization was followed by a 50~ps $NPT$ equilibration run at 298.15~
K and 1~bar using a timepstep of 0.25~fs.
We used the Berendsen thermostat and
barostat with $\tau_T$ = 1~ps and $\tau_P$ = 1~ps during equilibration.
Constant value of $4.5\cdot10^{-5}$~bar$^{-1}$ for the compressibility of the mixtures was
employed. In the case of pure methanol solvent, the compressibility
was taken to be $1.2\cdot10^{-4}$~bar$^{-1}$.
The V-rescale thermostat and Parrinello-Rahman
barostat with $\tau_T$ = 0.5~ps and $\tau_P$ = 2.0~ps
and the time step 2~fs were used during production runs.
Statistics for each mole solvent composition and various ions
concentration for any of the properties were collected over
several 10~ns $NPT$ runs, each starting from the last configuration of the
preceding run. The time extension for each series of calculations will be mentioned below
in the appropriate place but not less than 70~ns.

While exploring the composition changes, 
the total number of molecules 
was kept fixed at 3000. 
The composition is described by the mole fraction of methanol molecules $X_\text{m}$, $X_\text{m}=N_\text{m}/(N_\text{m}+N_\text{w})$.

\section{Results and Discussion}

\begin{figure}[!t]
	\begin{center}
		\includegraphics[width=8cm,clip]{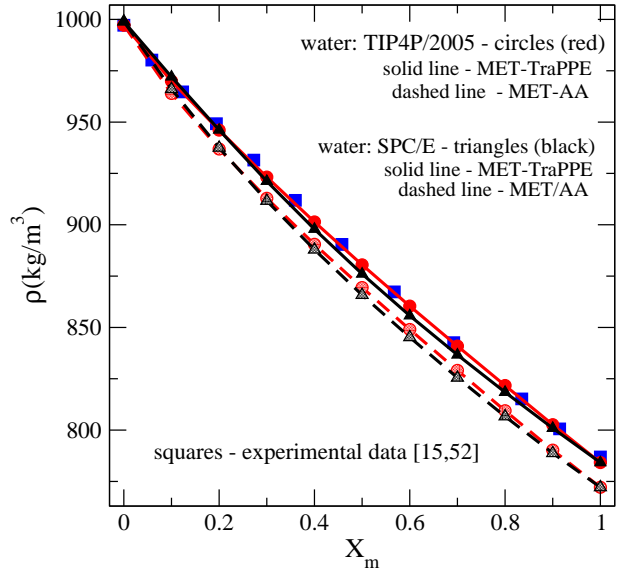}
	\end{center}
\vspace{-6mm}
	\caption{(Colour online) Composition dependence of water-methanol mixture density from the present $NPT$ 
		simulations of the TIP4P/2005-MET/TraPPE, TIP4P/2005-MET/AA, SPC/E-MET/TraPPE, and 
		SPC/E-MET/AA models, together with experimental data~\cite{mikhail,washbrun}. 
		The nomenclature of lines and symbols is given in the figure.}
	\protect
	\label{fig1}
\end{figure}

In order to make things clear from the very beginning, we study four models of water-methanol
liquid mixtures described by the following force fields: TIP4P/2005-MET/TraPPE,
TIP4P/2005-MET/AA, SPC/E-MET/TraPPE and SPC/E-MET/AA. In all cases, the Lorentz-Berthelot
combination rules are used to obtain cross interaction parameters.
The behavior of mixture density on composition, $X_\text{m}$,  is given in figure~\ref{fig1}.  
Our $NPT$ computer simulation results are supplemented by the experimental data~\cite{mikhail,washbrun}. 
It can be seen that the density is perfectly well described by the TIP4P/2005-MET/TraPPE
model. The SPC/E-MET/TraPPE combination of force fields yields only a bit worse
predictions in the interval of intermediate compositions. Two models, the TIP4P/2005-MET/AA
and SPC/E-MET/AA, are less accurate regarding the density dependence on composition, apparently
because the density of pure liquid methanol is not described accurately within the MET/AA model. 
These trends of behavior have been mentioned recently in~\cite{galicia1,galicia2}
exploring TIP4P/Ew water model~\cite{horn} combined with methanol models with a different 
degree of sophistication.
On the other hand, the density dependence on composition from the $NPT$ simulations 
by Soetens and Bopp
also agrees with the experimental data perfectly well~\cite{bopp2}. It is worth mentioning 
that the authors used well established BJH~\cite{bopp} and PHH~\cite{palinkas} 
flexible models for water and methanol, respectively.
The results presented in figures~2 and 3 of~\cite{jordan} are slightly off all the observations mentioned above concerning the density dependence on methanol 
molar fractions. Namely, in figure~2 
of that work, the liquid
density of pure methanol from TraPPE and OPLS-AA models at 300~K looks too low compared to the 
values reported in TraPPE database and by other authors. Moreover, the discrepancy between
the simulation predictions and experimental data given in figure~3 of~\cite{jordan} (the $y$-scale
of this figure seems to be not very appropriate to appreciate the total density) is too
big compared to other works. Apparently, trends of behavior of density from this work~\cite{jordan}
are affected by the number of particles, system size, method of calculations using 
a rather small liquid slab surrounded by vacuum and by simulation time.
To summarize the discussion of our figure~\ref{fig1} and the observations of other authors, it seems 
that the dependence of density on composition is one of the important properties, 
but not too demanding computationally, 
if the system size and time of simulations are properly chosen.

\begin{figure}[!t]
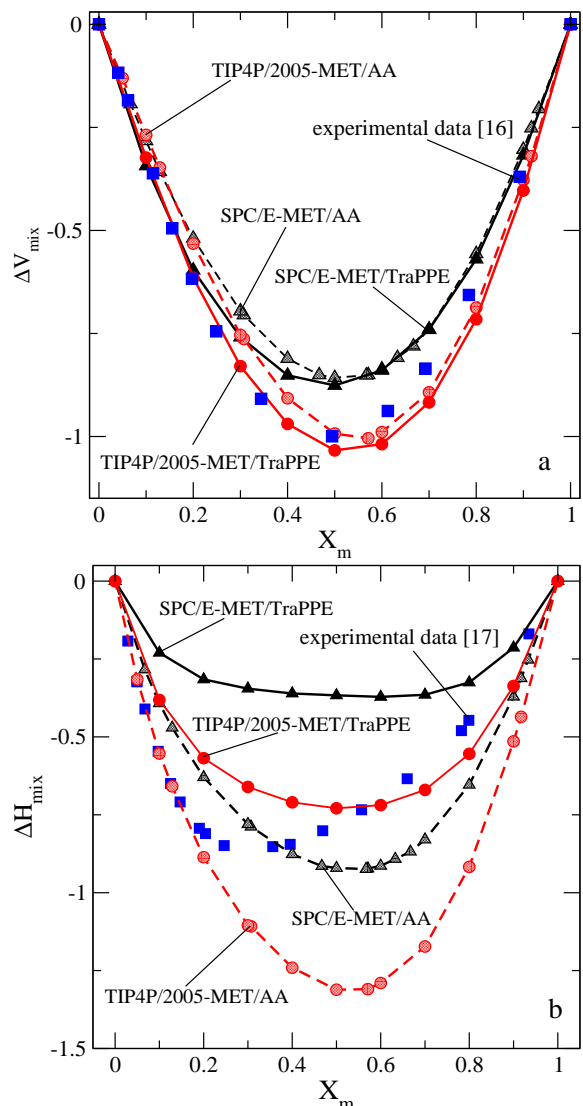

	\begin{center}
		\includegraphics[width=7.5cm,clip]{fig_delta_ve.eps}
		\includegraphics[width=7.5cm,clip]{fig_delta_entr.eps}
	\end{center}
\vspace{-6mm}
	\caption{(Colour online) Panel (a): Excess mixing volume of water-methanol mixtures on methanol molar fraction
		for different combinations of models for each species
		together with the experimental data from reference~\cite{mcglashan} (blue squares).
		Panel (b):  A comparison of computer simulation data for mixing entalphy with 
		experimental data from~\cite{lama} (blue squares). The nomenclature of lines and
		symbols is like in figure~\ref{fig1}.}
	%Both panels show TIP4P/2005 - MET/TraPPE,TIP4P/2005 - MET/AA, SPC/E - MET/TraPPE, 
	%and SPC/E - MET/AA models.}
	\protect
	\label{fig2}
\end{figure}

Next, in the spirit of previous works from our laboratory~\cite{galicia1,galicia2}  and 
of the development by Soetens and Bopp~\cite{bopp2}, we turn our attention to the mixing 
properties. The excess molar volume, $\Delta V_\text{mix}$, is defined as,
\begin{equation}
\label{eq1}
 \Delta V_\text{mix}=V_\text{mix}-(1-X_\text{m})V_\text{w}-X_\text{m}V_\text{m}\,,
\end{equation}
where $V_\text{mix}$ is the volume of the mixture at a certain composition, $V_\text{w}$ and $V_\text{m}$ refer
to the molar volumes of pure water and pure methanol, respectively. Our simulation results 
for $\Delta V_\text{mix}$ are shown in figure~\ref{fig2}~(a).
As in the case of density, cf. figure~\ref{fig1}, here we observe again that the TIP4P/2005-MET/TraPPE model
yields the best agreement with experimental results~\cite{lama}. The entire set of experimental points is
well reproduced, slight overestimation of the magnitude of $\Delta V_\text{mix}$ is observed in the
interval of intermediate compositions and when methanol content is higher than of water.
Nevertheless, the experimentally observed  minimum at $X_\text{m}=0.5$ 
is reproduced by simulations. If methanol is described
by all-atom model (MET/AA), the agreement between simulation results and experimental points
deteriorates, the minimum shifts to $X_\text{m}\approx 0.6$. Quite similar dependence 
of $\Delta V_\text{mix}$ on composition (to the TIP4P/2005-MET/AA force field) has been obtained 
from simulations of flexible BJH-PHH model, cf. figure~2 of reference~\cite{bopp2}. If one combines the SPC/E
water with each of two methanol models in question, the values for $\Delta V_\text{mix}$ in almost entire 
range of composition are underestimated, figure~\ref{fig2}~(a). This kind of inaccuracy has been observed 
previously and was discussed in detail in references~\cite{galicia2,nezbeda}.
Energetic aspects of mixing are given by the excess enthalpy of mixing, $\Delta H_\text{mix}$. It is
defined similarly to equation~\ref{eq1},
\begin{equation}
 \Delta H_\text{mix}=H_\text{mix}-(1-X_\text{m})H_\text{w}-X_\text{m}X_\text{m}\,,
\end{equation}
where $H_\text{mix}$ refers to the mixture entalphy whereas $H_\text{w}$ and $H_\text{m}$ describe the entalphy of each
species in pure state at the same temperature and pressure. From the comparison of the
simulation results in figure~\ref{fig2}~(b) with experimental data~\cite{lama}, we conclude that none of the employed
force field combinations reproduce the experimental trends perfectly well. Concerning the magnitude of
$\Delta H_\text{mix}$, the TIP4P/2005-MET/TraPPE and SPC/E-MET/AA models provide a better
description than the TIP4P/2005-MET/AA and SPC/E-MET/TraPPE. However, all the models
of this study predict the minimum of $\Delta H_\text{mix}$ at $X_\text{m}$ in the interval between 0.5 and 0.6
rather than the experiment that shows the corresponding minimum at $X_\text{m}=0.3$. Similar kind
of inaccuracy has been observed and discussed for a set of previously studied 
models~\cite{galicia1,galicia2}. Unfortunately, the excess entalphy of mixing from $NPT$ simulations
has not been presented in reference~\cite{bopp2} for the flexible BJH-PHH model to provide a critical
evaluation of the present modelling. Three data points previously reported for the
excess potential energy from the NVE simulations of the combination of flexible models, 
see table~1 of reference~\cite{old-palinkas}, do not provide a definite answer in this respect, unfortunately.
 
The microscopic structure of water-methanol mixtures has been discussed in terms of
evolution of the pair distribution functions on composition in many occasions. However,
quantitative insights about the changes of structure on $X_\text{m}$ are usually described in terms
of the first coordination numbers of the species and of the number of hydrogen bonds. Here,
in order to avoid unnecessary repetition, we adopt a similar point of view.
The first coordination numbers of the species are defined from the running coordination numbers
as common,
\begin{equation}
n_{ij}(r)=4\piup\rho_j \int _0 ^r  g_{ij}(R)R^2 \rd R\,,
\end{equation}
where the upper limit of integration in this equation is taken to correspond to the first mi\-ni\-mum 
$r_\text{min}$ of the pair distribution function $g_{ij}(r)$ of the species $i$ and $j$.
A detailed description of the behavior of this property, 
for a set of models related to the present study, was given in~\cite{galicia1,galicia2,galicia3}.
Very recent reports also concern the evolution of the first coordination numbers, 
see, e.g., figure~7 of reference~\cite{bopp2} and figures~24 and 25 of reference~\cite{jordan}.

In order to evaluate the average number of H-bonds between molecules, the corresponding utility
of GROMACS software is applied with default distance --- angle criterion. However, for distance
cutoff we used the first minimum of the corresponding pair distribution function,
see also discussion of this issue in references~\cite{galicia1,galicia2,galicia3}. 
Averaging is performed over a piece of or the entire simulated trajectory.

\begin{figure}[!b]
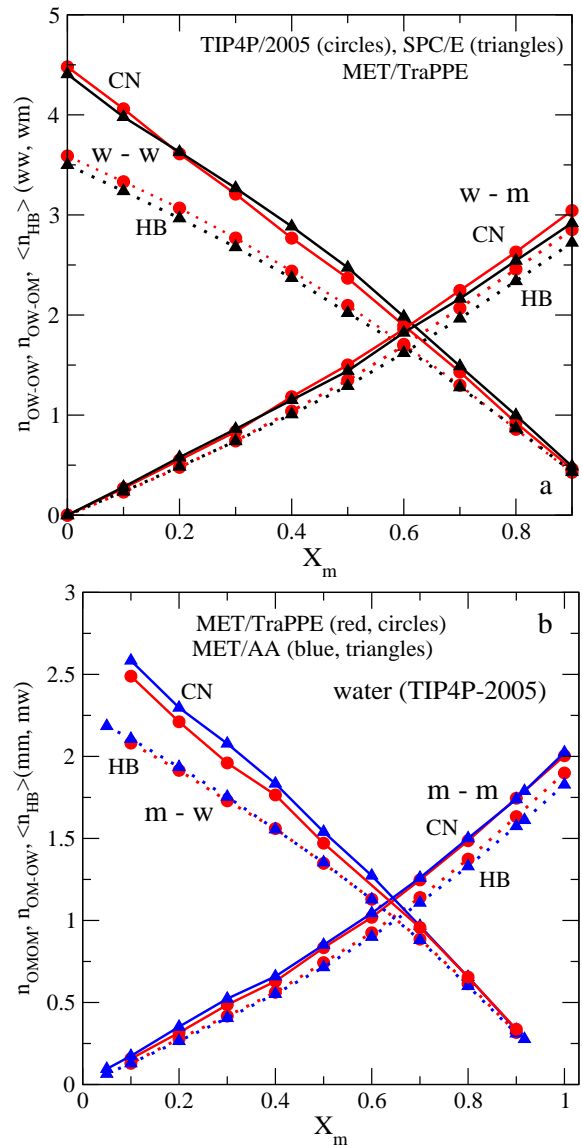

	\vspace{-2mm}
	\begin{center}
		\includegraphics[width=7.5cm,clip]{coord_wat.eps}
		\includegraphics[width=7.5cm,clip]{coord_methanol.eps}
	\end{center}
\vspace{-6mm}
	\caption{(Colour online) The dependence of the first coordination number of oxygens for
		each species (CN) and of average number 
		of H-bonds (HB) on composition for models as in figure~\ref{fig1} and figure~\ref{fig2}. The
		coordination numbers are given by solid lines and symbols, whereas 
		the dotted lines refer to the numbers of hydrogen bonds.
		Normalization of the average number of H-bonds is performed  per number of molecules 
		of the first species of the notation:  w-w, w-m and m-m and m-w. Panel (a) is for
		w-w and w-m where the panel (b) os for m-m and m-w.
		The nomenclature of lines and symbols is given in the figure.}
	\protect
\label{fig3}
\end{figure}

Changes of $n_{ij}=n_{ij}(r_\text{min})$ on $X_\text{m}$ for oxygens that belong to water or methanol,
OW-OW and OW-OM, and resulting from our simulations, are shown in figure~\ref{fig3}(a). 
The coordination number $n_\text{OW-OW}$ monotonously,
almost linearly,
decreases with an increasing methanol mole fraction starting from the value $\approx 4.5$ for pure water.
On the other hand, $n_\text{OW-OM}$ increases up to $\approx 3$  while approaching $X_\text{m}=1$.
Two combinations of force fields, TIP4P-2005-MET/TraPPE and SPC/E-MET/TraPPE, yield
almost the same results. The crossing point between $n_\text{OW-OW}(X_\text{m})$ and $n_\text{OW-OM}(X_\text{m})$
describing ``inversion'' in the type of predominant neighbors around water molecule 
occurs at $X_\text{m} \approx 0.65$. The crossing point at the same composition characterizes 
the BJH-PHH combination of flexible models for two species, figure~7 of~\cite{bopp2}. 
On the other hand, evolution of the average number of hydrogen bonds $\langle n_\text{HB} \rangle$ per water molecule
on the methanol mole fraction from our simulations is given in figure~\ref{fig3}~(a) as well.
Actually the dependences of $\langle n_\text{HB} \rangle$ on $X_\text{m}$ follow the trends of behavior of the first 
coordination numbers. The crossing point describing the predominant number of H-bonds 
between two kinds of molecular species is observed at $X_\text{m} \approx 0.65$. Again, 
two water models (TIP4P/2005 and SPC/E), if combined with MET/TraPPE, yield very similar
predictions concerning the trends of behavior of $\langle n_\text{HB} \rangle$ on $X_\text{m}$.   
%%%  make figures with MET/AA to be sure that it yields the same!!!  %%%%%%%%%%%%%%%
The ``inversion'' or crossover point in $n_\text{OW-OW}(X_\text{m})$ and $n_\text{OW-OM}(X_\text{m})$ or $\langle n_\text{HB} \rangle$
is located close to the minimum of the excess mixing volume, cf. figure~2~(a), as it was
discussed in~\cite{galicia2}. Finally, we would like to mention that the approximation of
average hydrogen bond numbers by the corresponding coordination numbers is not very
appropriate in the case OW-OW, see figure~\ref{fig3}~(a), in contrast to what was claimed in~\cite{jordan}.

Concerning the behavior of methanol coordination number and the average number of hydrogen bonds
in a mixture with varying composition, figure~\ref{fig3}~(b), we would like to mention the following trends.
We explored two methanol models, the MET/TraPPE and MET/AA, combined with TIP4P-2005 water
model. The models yield a similar behavior. In pure methanol, the coordination number is
around three and the average number of hydrogen bonds between methanol molecules is only
slightly less. The methanol coordination smoothly decreases with a decreasing $X_\text{m}$. On the other hand,
the cross coordination number OM-OW increases with a decreasing $X_\text{m}$, the corresponding
hydrogen bonds number follows this behavior. The ``inversion'' of the composition of the surrounding of
a methanol molecule on average occurs at $X_\text{m} \approx 0.65$. In the region of low methanol mole
fractions, methanol molecule incorporates the hydrogen bonded structure of water and on average forms a bit more than two hydrogen bonds with water molecules.

An excellent detailed analysis of the coordination numbers and hydrogen bonds network 
topology was 
performed in reference~\cite{chihaia}, though the energetic definition of hydrogen bonds was
used in that work. Nevertheless, all trends of behavior deduced from geometric
criterion for hydrogen bonding qualitatively agree with the predictions described by using
energetic definition.

%%%%%%%%  MICROHETEROGENEITY  add!!!!!!!!!!!!!!!!  experiment Takamuku! %%%%%

One of the properties representing a quite demanding test of the employed force field
is the surface tension that describes the changes of surface free energy upon 
changing the surface area. 
Therefore, we have undertaken additional calculations focused
on the evaluation of surface tension for a set of models under study. Relevant
experimental results are available~\cite{vazquez}. Moreover, this property
has been explored in detail in several computer simulation 
studies of water-methanol system, see, e.g.,~\cite{jordan,matsumoto,chang}.

The simulations aiming at surface tension calculations at each point of composition
axis in terms of $X_\text{m}$, have been performed by using the final configuration of particles 
in the box from the $NPT$ run.
Next, the box edge along $z$-axis was elongated by a factor of 3, generating a box with
liquid slab and two liquid-mixture -- vacuum interfaces in the $x-y$ plane,
in close similarity to the procedure used in reference~\cite{fischer}.
The total number of particles, $3\times 10^3$, seem to be reasonable as it provides the area of 
the $x-y$ face of the liquid slab large enough to avoid size effects. 
The elongation of the liquid slab along $z$-axis is sufficient as well.
To be more specific, we would like just to mention that the box dimensions 
were $4.63\times 4.63 \times 13.89$~nm$^3$ for $X_\text{m}=0.1$ up to $5.75 \times 5.75 \times 17.25 $~nm$^3$ for $X_\text{m}=0.9$,
for the TIP4P/2005-MET/TraPPE model calculations. These numbers favourably compare 
with the cut-off distance for non-bonded interactions $1.4$~nm.  
The executable molecular dynamics file was modified by
deleting fixed pressure condition, just the  $V$-rescale thermostatting with the same
parameters as in the $NPT$ runs has been preserved. Other corrections have not been used.

The values for the surface tension, $\gamma$, follow from the combination
of the time averages for the components of the pressure tensor,
\begin{equation}
\gamma = L_z \Big\langle\Big[P_{zz} - \frac {1}{2} \left(P_{xx}+P_{yy}\right)\Big]\Big\rangle \Big/ 2\,,
\end{equation}
where $P_{ij}$ ($i,j = x,y,z$) are the components of the pressure tensor,
and $\langle ... \rangle$ denotes the time average.
We performed a set of $NVT$ runs, not less than 5--6, each with the time duration of $10$~ns,
and obtained the results for $\gamma$ making the block average.

\begin{figure}[!b]
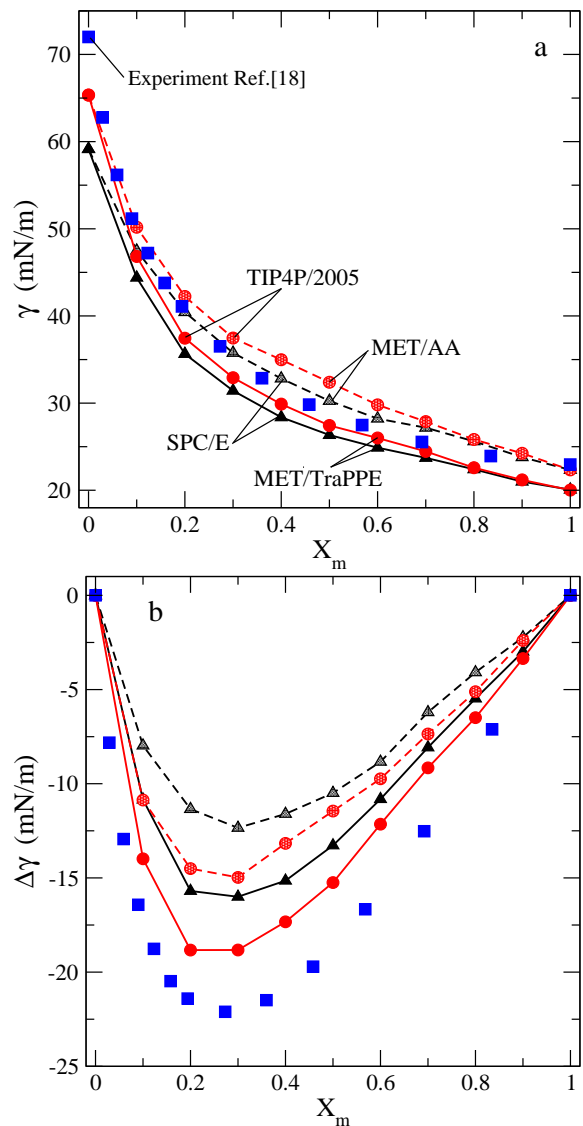

	\begin{center}
		\includegraphics[width=7.5cm,clip]{sufr_tens_a.eps}
		\includegraphics[width=7.5cm,clip]{sufr_tens_b.eps}
	\end{center}
\vspace{-6mm}
	\caption{(Colour online) Panel (a): Surface tension of water-methanol mixtures on methanol molar fraction
		as compared with experimental data from reference~\cite{vazquez} (blue squares).
		Panel (b): Excess mixing surface tension on composition.
		Both panels refer to the same models as in previous figures.}
	\protect
	\label{fig4}
\end{figure}

Our results for the surface tension and the excess mixing surface tension are given
in two panels of figure~\ref{fig4}. It is known that the TIP4P/2005 model leads to a better 
prediction for the surface tension of pure water in comparison with SPC/E~\cite{vega-pccp}.
This is also documented in figure~\ref{fig4}~(a).
Our result for pure methanol within MET/TraPPE model agrees perfectly well with reference~\cite{biscay}.
On the other hand, the application of the same procedure with MET/AA model yields a
slightly higher value for pure methanol, very close to the experimental
result, figure~\ref{fig4}~(a). Small discrepancy of our result with the value given in the
supplementary material to~\cite{fischer} can be attributed to slightly different 
technical details of simulations, e.g., cut-off, corrections to the
electrostatics for a slab system, etc.

Overall  trends of behavior of $\gamma (X_\text{m})$ in the entire composition interval
are qualitatively correctly reproduced by all four combinations of the force fields.
A better description at low values of $X_\text{m}$ is reached if the TIP4P/2005 model is
invloved. For high values of $X_\text{m}$, the application of the MET/AA model leads to a
better agreement with the experimental data. The values calculated by us are lower 
than the experimental data~\cite{vazquez} at all $X_\text{m}$, due to the defficiency of performance of 
models for each component, if the TIP4P/2005 and MET/TraPPE are used. With the 
MET/AA model of methanol, the surface tension values are higher than the experimental
points at intermediate and high methanol mole fractions. Interestingly, the surface
tension substantially decreases even if a small amount of methanol is added to water.
This behavior can be attributed to the tendency for methanol to be located close to
the interface with vacuum. Our findings are in agreement with the results reported
in~\cite{jordan} for the same system but with a smaller number of particles.

If one focuses on the deviation
of the surface tension from ideal mixing behavior, figure~\ref{fig4}~(b), the TIP4P/2005-MET/TraPPE
model is the closest to the experimental predictions. The present
satisfactory modelling can be seemingly used with confidence to other more complex systems.
It is worth mentioning that experimental results predict a maximum absolute value of
deviation from ideality at $X_\text{m} \approx 0.3$ and computer simulation results
reproduce this behavior. This particular composition actually coincides
with the maximum deviation of mixing entalphy, cf. figure~\ref{fig2}~(b), that has not been
predicted by the models in question. Seemingly, the 
local composition fluctuations with specific orientations of molecules (yielding
nonideality of surface tension) are missing in the bulk phase to provide a 
correct behavior of nonideality of entalphy.
At present, it is difficult to offer a recipe how to improve the performance of nonpolarizable 
rigid models for surface tension and for nonideality of entalphy, though. 

\begin{figure}[!b]
	\begin{center}
		\includegraphics[width=7.5cm,clip]{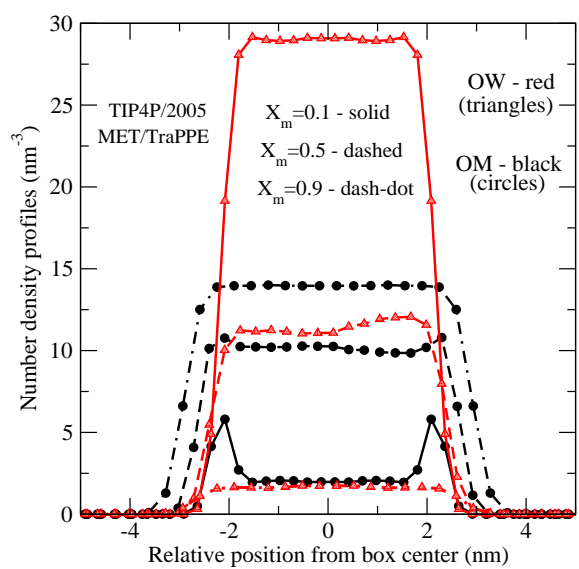}
	\end{center}
	\caption{(Colour online) Examples of the number density profiles of water and methanol oxygens for mixtures 
		(TIP4P/2005-MET/TraPPE model) at different composition.}
	\protect
	\label{fig5}
\end{figure}

In order to explore how the particles of different species are distributed in the box when
$X_\text{m}$ changes, we have plotted the number density profiles of water and methanol 
oxygens in the entire slab plus vacuum system in  figure~\ref{fig5}. At a low value of $X_\text{m}$,
see $X_\text{m}=0.1$ as an example, water is uniformly distributed in the slab. Methanol species
are distributed uniformly in the inner part of the slab, though we observe a high
maximum for methanol oxygens distribution  at each liquid slab --- vacuum interface. 
At intermediate composition, $X_\text{m}=0.5$, both species are almost uniformly distributed
inside the liquid slab, but the maxima of $\text O_\text m$ observed for $X_\text{m}=0.1$ disappear, in expence 
of the growing amount of methanol oxygens in the part of each interface exposed to vacuum. 
The two profiles describing the intermediate composition, $X_\text{m}=0.5$, are not perfectly
symmetric with respect to the box center. This situation occurs very rarely
but can have consequences, if one pretends to obtain the precise density of species 
from this kind of procedure. Specifically, in reference~\cite{jordan}
the authors failed to estimate the liquid density both for methanol and water at
$X_\text{m}=0.3$.  
Finally, at $X_\text{m}=0.9$ in figure~\ref{fig5}, it can be seen that the interface predominantly contains methanol molecules 
whereas water molecules ``hide'' in the inner part of the liquid slab. Only a small
fraction of water molecules penetrates the interfacial region.

\begin{figure}[!b]
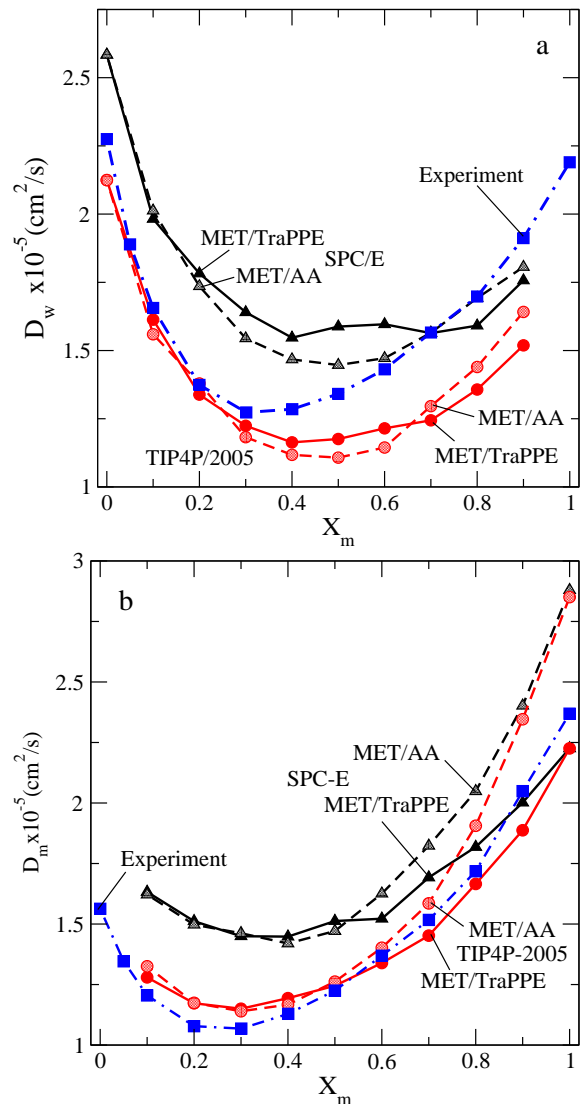

	\begin{center}
		\includegraphics[width=7.5cm,clip]{dif_water.eps}
		\includegraphics[width=7.5cm,clip]{dif_methanol.eps}
	\end{center}
\vspace{-6mm}
	\caption{(Color online) Self-diffusion coefficients of water, D$_\text{w}$,  and of 
		methanol, D$_\text{m}$, in  water-methanol mixture on composition [panels (a) and (b), respectively] 
		and experimental data~\cite{derlacki}.
		The nomenclature of lines and symbols is given in the figure.}
	\protect
\label{fig6}
\end{figure}

The final part of the manuscript is concerned with the description of the 
self-diffusion coefficients of species and changes of the dielectric constant
with composition of water-methanol mixtures.
The self-diffusion coefficients of water and methanol were calculated 
 from the mean-square displacement (MSD) of a particle via Einstein relation,
\begin{equation}
D_i =\frac{1}{6} \lim_{t \rightarrow \infty} \frac{\rd}{\rd t} \langle\vert {\bf 
r}_i(\tau+t)-{\bf r}_i(\tau)\vert ^2\rangle\,,
\end{equation}
where $i$ refers to water or methanol and $\tau$ denotes the time origin. 
Default settings of GROMACS were used for the separation of the time origins. 
Moreover, the  fitting interval (from 10\% to 50\% of the 
analyzed trajectory) has been used to calculate $D_\text{m}$ and $D_\text{w}$.
Moreover, a special care has been taken to fitting for the cases with a small 
number of particles on the extremes along the $X_\text{m}$ axis. 
A set of our results is given in  figure~\ref{fig6}. 

The results for $D_\text{w}$, are qualitatively correct, 
the  TIP4P/2005 model slightly underestimates $D_\text{w}$, it yields 2.1 for pure
water, whereas the SPC/E overestimates $D_\text{w}$, it yields $\approx 2.6$, see figure~\ref{fig6}~(a).
Concerning the self-diffusion coefficient of pure methanol, it 
can be seen that MET/TraPPE underestimates $D_\text{m}$, it gives $\approx 2.25$, whereas
the all atom model, MET/AA, substantially overestimates $D_\text m$, figure~\ref{fig6}~(b). This tendency
has been already discussed in reference~\cite{galicia3}. These inaccuracies prohibit
quantitatively correct predictions for the composition changes of the self-diffusion
coefficients of species. 

Namely, the dependence $D_\text{w}(X_\text{m})$ is qualitatively correct. The TIP4P/2005, if combined
with \linebreak MET/TraPPE or with MET/AA methanol model, leads to a very good prediction of $D_\text{w}$
up to $X_\text{m} \approx 0.3$, for higher values of $X_\text{m}$ simulation data deviate from 
experimental results~\cite{derlacki}. 
The miminum value for $D_\text{w}$ from simulations is in the interval 
$X_\text{m}$ between 0.4 and 0.5 whereas the experiment predicts this minimum at 0.3.
The minimum value of $D_\text{w}$ from simulations along the composition axis coincides with the minimum of 
excess mixing volume at $X_\text{m}=0.5$. At high values of $X_\text{m}$, the growth of $D_\text{w}$ is similar 
in simulations and in experiment~\cite{derlacki}. 
If water is descibed in the framework of the SPC/E
model, the resuts for $D_\text{w}(X_\text{m})$ substantially overestimate this self-diffusion 
coefficient in a wide range of composition. Solely at high values of $X_\text{m}$, the agreement
with experimental values becomes more acceptable. It is worth mentioning that we
were unable to get a better shape of $D_\text{w}(X_\text{m})$ within the SPC/E-MET/TraPPE model. 
Apparently, there are two minima, one at $X_\text{m}=0.4$ and anoter at $X_\text{m}=0.7$. It is
difficult to establish (without additional exploration of various properties) if
this behavior is related to the clustering of species at local scale as it has
been discussed in the experimental study~\cite{takamuku}. Here, we would like
just to mention that similar evolution of $D_\text{w}$ on $X_\text{m}$ was reported in
our recent work~\cite{galicia2} in the study of SPC/E model combined by OPLS/UA model 
of methanol~\cite{jorgensen}.  

Concerning the trends of behavior of the self-diffusion of methanol species in mixtures
of different composition $D_\text{m}(X_\text{m})$, we would like to emphasize the following. 
The best combination of force fields is provided by the TIP4P/2005-MET/TraPPE model.
It describes the changes of the function $D_\text{m}(X_\text{m})$ pretty well in the entire composition 
range, figure~\ref{fig6}~(b). The minimum value of $D_\text m$ is described at a correct place, $X_\text{m} \approx 0.3$.
All other combinations of the force fields exhibit defficiencies either due to the water model
like SPC/E or due to all-atom modelling of methanol at MET/AA level.  
An overall most satisfactory picture of the dependence of the self-diffusion coefficients 
of two species, therefore, results if the TIP4P/2005-MET/TraPPE model is used.
We believe that alternative calculations of the self-diffusion coefficients by 
applying velocity autocorrelation functions should lead to similar conclusions.

Our final concern is the evolution of the dielectric constant with composition.
The long-range, asymptotic behavior of correlations between molecules possessing a
permanent dipole moment is determined by the dielectric 
constant, $\varepsilon$.
Usually,  long molecular dynamics runs are necessary to obtain
reasonable values for $\varepsilon$,
because it is calculated from the time-average of the fluctuations of the total
dipole moment of the system~\cite{martin},
\begin{equation}
\varepsilon=1+\frac{4\piup}{3k_\text{B}TV}\big(\langle\bf M^2\rangle-\langle\bf M\rangle^2\big),
\end{equation}
where $k_\text B$ is the Boltzmann constant and V is the volume of the simulation box.

\begin{figure}[!t]
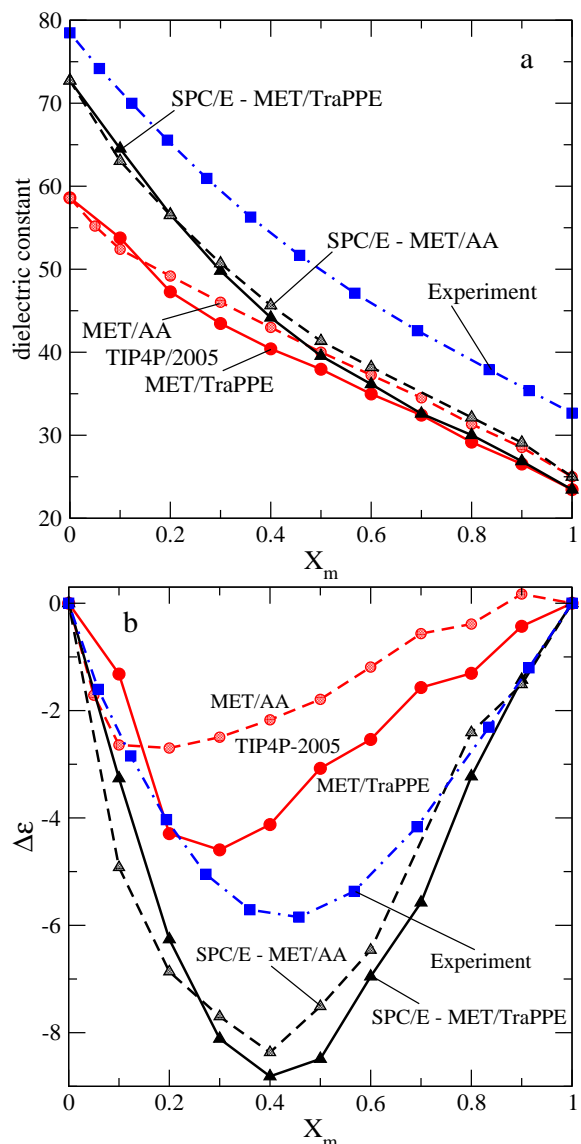

	\begin{center}
		\includegraphics[width=7.5cm,clip]{epsilon.eps}
		\includegraphics[width=7.5cm,clip]{delta_epsilon.eps}
	\end{center}
\vspace{-6mm}
	\caption{(Colour online) Panel (a): Dielectric constant of water-methanol mixtures on 
		methanol molar fraction and  experimental data from reference~\cite{albright} (blue squares).
		Panel (b): Excess  dielectric constant on composition.
		The experimental data are from~\cite{albright} (blue squares).
		The nomenclature of lines and symbols is given in the figure.}
	\protect
	\label{fig7}
\vspace{-4mm}
\end{figure}

%In this work we have taken
%care that each of the runs is of sufficient length, not less than $60~ns$ in all cases.
The lines from our $NPT$ simulations of $\varepsilon$ are shown in figure~\ref{fig7}~(a).
An overall behaviour of $\varepsilon(X_\text{m})$ is that it decreases
starting from a high value for pure water
to a lower value corresponding to pure methanol with an increasing $X_\text{m}$.
As it follows from the comparison of the simulation results and experimental 
data~\cite{albright},
all four combinations of the force fields substantially underestimate the values 
for $\varepsilon$ in the entire composition range. The reason is that the
dielectric constant for two constituents, water and methanol, is
essentially underestimated.
It is highly probable to improve the dependence of the static dielectric constant
on composition by applying the model for water specifically parametrized to reproduce
the dielectric constant for water~\cite{alejandre}. Simultaneously, it would require
parametrization of the force field for methanol, see, e.g.,~\cite{dominguez}.
However, common experience is that  parametrization of a single property leads to worse 
predictions for other properties. Therefore, this issue requires additional
computation efforts. 

On the other hand, a sensitive test is provided by comparison of the excess
dielectric constant,
$\Delta \varepsilon^\text{mix} = \varepsilon_\text{m}- [X_\text{m}\varepsilon_\text{m}+(1-X_\text{m})\varepsilon_\text{w}]$,
with the experimental predictions~\cite{albright}.
Experimental points indicate a negative deviation from
ideality in the entire composition range, figure~\ref{fig6}~(b).
Maximal (negative) deviation from the ideal type of behaviour
reported from the experimental measurements is at $X_\text{m} \approx  0.45$.
The simulations results  reproduce the position of a minimum approximately; namely,
the minimum is in the interval 0.15--0.4, dependent on the combination of the force fields.
These trends are in accordance with observations concerning the deviations from ideality
of thermodynamic properties.
The magnitude of the excess static dielectric constant
is overestimated if the SPC/E model for water is used and is
underestimated if TIP4P/2005 is involved.
The TIP4P/2005-MET/TraPPE model is the most close to the experimental data combination of models. The excess dielectric
constant curve as function of chemical composition of the mixture
can be related to the excess refractive index measurements, see, e.g.,~\cite{gofurov}.
Consequently, a complementary comparisons with experiments would be desirable in future
work.

\section{Summary and conclusions}

This work has been principally inspired by the necessity to evaluate
various properties of water-methanol mixtures using a specific set of 
models for each component. The choice of the force fields of this study 
permits extensions necessary to describe ionic solutions with such combined solvents
in the spirit of very recent contributions from the research
laboratory of C. Vega~\cite{benavides1,madrid} concerning 
aqueous NaCl solutions with a novel force field for ions.

In the present work, extensive $NPT$  molecular dynamics simulations were conducted to study
thermodynamic, dynamic and structural properties of water-methanol
mixtures. Two different water and two methanol nonpolarizable models 
were combined and simulated with the purpose of testing their predictions
for an ample set of properties in the entire range of compositions. 
Comparisons with the available experimental data were performed.
Considering the scope of the models, the predictions obtained for
the mixtures appear to be qualitatively correct, particular properties
were a bit better described at low methanol compositions, i.e., for
water-rich compositions. However, as a general trend, it is observed
that the best predictions are given with the
water (TIP4P/2005)-methanol(TraPPE) mixtures since these models reasonably well predict several properties of the pure components.
The results indicate that a good agreement with laboratory experiments
could be obtained when both force fields, of the two components in the mixture,
are good. In fact, the dielectric constant is not well predicted by any of the
simulated mixtures since none of the selected models (water and methanol)
predicts correctly that property.

\vspace{-3mm}
\section*{Acknowledgements}
M. Cruz and O.P. are grateful to M. Aguilar for technical support of this work
at the Institute of Chemistry of the UNAM. M. Cruz acknowledges support of CONACyT 
of Mexico for Ph.D. scholarship.

\appendix
%----------------------------------------------------------------------------------------------------------------------

\ukrainianpart 

\title{Моделювання методом молекулярної динаміки властивостей  сумішей вода - метанол.  Вплив силових полів} 

\author{М. Круз Санчес\refaddr{label1}, Г. Домінгес\refaddr{label2},
	О. Пізіо\refaddr{label1}}

\addresses{
	\addr{label1} Інститут хімії,  Нацiональний автономний унiверситет м. Мехiко, Мехiко, Мексика
	\addr{label2} Інститут матеріалознавства,  Нацiональний автономний унiверситет м. Мехiко, Мехiко, Мексика
}

\makeukrtitle

\begin{abstract}
	Моделювання методом молекулярної динаміки в ізотермічно-ізобаричному ансамблі застосовано до
	дослідження  мікроскопічної структури та деяких властивостей  рідкої суміші вода-метанол.
	Моделі води  TIP4P/2005 і SPC/E поєднано з  моделлю об'єднананих атомів  TraPPE і моделлю силових полів всіх атомів для метанолу. 
	Основною метою даної роботи є дати якісну оцінку передбаченням різних комбінацій модельних силових полів стосовно
	концентраційних залежностей основних властивостей системи. Зокрема, ми дослідили вплив концентрації на густину,
	надлишковий молярний об'єм і надлишкову ентропію, а також на поверхневий натяг і статичну діелектричну сталу.
	Крім цього, описано структурні властивості на мові координаційних чисел і середнього числа водневих зв'язків
	між молекулами компонентів суміші. Нарешті, здійснено оцінку концентраційної залежності коефіцієнтів самодифузії
	компонентів. Усі теоретичні передбачення перевірено по відношенню до експериментальних даних.
	
	\keywords  суміші вода-метанол, властивості змішування, поверхневий натяг, моделювання методом молекулярної динаміки
\end{abstract}

%fig1

%fig2

%fig3

%fig4

%fig5

%fig6

%fig7

\end{document}